\documentclass[aps,prd,twocolumn,nofootinbib,superscriptaddress]{revtex4-1} 
\usepackage{fullpage, amsmath, xparse, amssymb, mathtools, graphicx, braket,xcolor,verbatim}
\usepackage{natbib}
\usepackage[colorlinks=true,linkcolor=blue,citecolor=blue]{hyperref}

\usepackage{multirow}
\usepackage{cancel}

\newcommand\lsim{\mathrel{\rlap{\lower4pt\hbox{\hskip1pt$\sim$}}
        \raise1pt\hbox{$<$}}}
\newcommand\gsim{\mathrel{\rlap{\lower4pt\hbox{\hskip1pt$\sim$}}
        \raise1pt\hbox{$>$}}}

\begin{document}
\title{Caustic fringes for wave dark matter}

\author{Andrew Eberhardt}
\thanks{Kavli IPMU Fellow}
\email{andrew.eberhardt@ipmu.jp}
\affiliation{Kavli Institute for the Physics and Mathematics of the Universe (WPI), UTIAS, The University of Tokyo, Chiba 277-8583, Japan}
\author{Lam Hui}
\email{lh399@columbia.edu}
\affiliation{Physics Department and Center for Theoretical Physics,
  Columbia University, New York, NY 10027, USA}
\email{lh399@columbia.edu}




\begin{abstract}
Wave dark matter is composed of particles sufficiently light that their de Broglie wavelength
exceeds the average inter-particle separation. A typical wave dark matter halo
exhibits granular substructures due to wave interference. In this paper, we explore the wave interference
effects around caustics. These are locations of formally divergent
density in cold collisionless systems.
Examples include splashback in galaxy clusters, and tidal shells in
merging galaxies, where the pile-up of dark matter close to apogee gives rise to caustics.
We show that wave interference modifies the density profile in the vicinity of the caustics,
giving rise to a fringe pattern well-described by the Airy function.
This follows from approximating the gravitational potential as linear close to apogee.
This prediction is verified in a series of numerical simulations in which the gravitational potential is computed exactly.
We provide a formula expressing the fringe separation in terms of the wave dark matter mass and halo parameters, which is useful for interpreting and stacking data.
The fringe separation near caustics can be significantly
larger than the naive de-Broglie scale
(the latter set by the system's velocity dispersion). This opens up the possibility
of detecting caustic fringes for a wide range of wave dark matter masses.
\end{abstract}

\maketitle

\section{Introduction}

One of the outstanding questions in cosmology concerns the nature of dark matter.
Is it particle-like, composed of a weakly interacting massive particle (WIMP) for instance,
or wave-like, composed of an axion or axion-like-particle?
The dividing line is a particle mass of about $10$ electron-volts (eV), below which the dark matter
effectively behaves as a collection of waves, with the
de Broglie wavelength exceeding the typical inter-particle separation
in a galaxy like our own. While the wave phenomenon of interest in this paper
applies to any wave dark matter less than $10$ eV, we will be particularly interested in the
ultra-light end of the spectrum for which there could be astrophysically observable consequences , i.e. a dark matter mass $m$ less than about $10^{-19}$ eV,
a possibility often known as fuzzy dark matter \cite{Hu2000}.
For recent reviews on fuzzy dark matter, or wave dark matter more generally,
see \cite{Ferreira2021,Hui2021,Eberhardt2025Review}.

An important implication of wave dark matter is the existence of granular
substructures or quasi-particles inside galaxy halos \cite{Schive2014,Hui:2016ltb}. These have been extensively studied in the literature in the context of the heating of stellar dispersions \cite{Hui:2016ltb, Amorisco:2018dcn, Bar_Or_2019, Church2019, dalal2022, DuttaChowdhury2023, teodori2025, Eberhardt2025_heating}, gravitational lensing \cite{Chan_2020, Hui:2020hbq, Liu2024, Broadhurst:2024ggk, Palencia:2025wjw, eberhardt2025, Amruth2023}, multiple ultralight fields \cite{Gosenca2023}, higher spin fields \cite{Amin2022}, quantum corrections \cite{Eberhardt2023}, 
filaments \cite{Zimmermann:2024vng}, astrometry \cite{Kim2024_astro,Dror:2024con}, and pulsar timing \cite{Kim2024, Eberhardt2024, Boddy:2025oxn, luu2024, Liu2023, xue2024} among others. 
We can think of a halo as consisting of a set of waves with a spread of momentum, given
by $m \sigma$ where $\sigma$ is the velocity dispersion of the halo.
Assuming the waves have random phases, it is straightforward to derive the
power spectrum of density fluctuations, for which the only relevant scale is 
the de Broglie wavelength $\lambda_{\rm dB} \equiv \hbar / m\sigma$.
(Henceforth, we reserve the symbol $\lambda_{\rm dB}$ for the de Broglie scale computed
using the average velocity dispersion $\sigma$.)
Essentially, wave interference creates granules of the size of $\lambda_{\rm dB}$, and
the density fluctuates by order unity from one granule to another.
A derivation of the density fluctuation statistics, and the verification thereof in
numerical simulations, can be found in
\cite{Hui:2020hbq,Yavetz:2021pbc}. See also \cite{Derevianko:2016vpm,Foster:2017hbq,Centers:2019dyn}.

We should expect substructures beyond such average de Broglie-sized granules.
Consider a shell of materials expanding outward through a galactic halo, the gravitational
pull of the halo will slow down those materials. They will tend to pile up at apogee, creating
a density spike. If the dark matter were particle-like, this is what one would call a caustic.
\footnote{We will be more careful in distinguishing between caustic and apogee later.}
If the dark matter were wave-like, we expect interesting wave interference effects around such a
density spike. In particular, because the underlying phase space sheet turns around near apogee (resulting in velocity separations smaller than the halo velocity dispersion),
one might expect the corresponding interference pattern to have a scale larger
than $\lambda_{\rm dB}$ (which is set by the average halo velocity dispersion).
This is indeed what we are going to find.

The phenomenon of density spikes around apogee has been observed in
a variety of contexts.
In the halo formation process from a quasi-spherical collapse,
splashback refers to the point at which the outer shell of materials first reaches
apogee after turnaround.
This has been observed in simulations and observations
of clusters \cite{More2016,Chang2018},
with the splashback radius identified by a kink in the cluster density profile---
the density has a precipitous drop beyond that shell at apogee
(see also \cite{Diemer2014,More2015,Vogelsberger2011,delpopolo2022}).
A second setting in which caustics are found is in galaxies which have
undergone mergers: the tidal debris expand outward, forming tidal shells
around apogee. Such tidal shells have been observed in both simulations and
observations
\cite{Malin1983,Sanderson2013,2015MNRAS.454.2472H,Pop2018,Kado-Fong2018,DongPaez2022}. 
Our goal is explore the predictions of wave dark matter for settings like these.

Our work builds on a very interesting paper by Gough and Uhlemann \cite{Gough2022}.
Their work focuses on interference behavior around caustics, for free particles or waves.
We extend their analysis to that of particles or waves moving in a linear potential,
as appropriate for modeling the slow-down close to apogee.
Earlier work on caustics in a galactic halo, emphasizing particles rather than waves, can be found in
\cite{Sikivie:1997ng,Duffy:2008dk,Chakrabarty:2018gdg}.

The paper is organized as follows. In Section \ref{AiryAll}, we
discuss how the Airy function approximately describes the behavior
of waves in the vicinity of a caustic, and how it takes a particularly
simple form if the caustic happens to be close to apogee.
We then test this analytic understanding, based on a linear potential,
against numerical simulations of the full Schr\"odinger-Poisson system
in Section \ref{NS}. The simulations are done in both 1D and 3D (one and three
spatial dimensions), and include 3D cases without spherical symmetry.
In parallel, we also present the results of the corresponding N-body
simulations which will allow us to compare and contrast the behavior
of particles versus waves.
We conclude in Section \ref{Discuss} with a discussion of the observational prospects and future work to be done.
We emphasize the scaling of fringe separation with acceleration, deduced in
Section \ref{AiryAll}, would be useful if stacking is required to pull the signal
out of noisy data. In Appendix \ref{HJ}, we provide a derivation of
the wave behavior around a caustic in a linear potential.
In Appendix \ref{2D}, we give a derivation of the Airy solution in
  2D.

A word on our terminology: we use the term {\it caustic fringes} to describe the interference pattern of interest. {\it Caustic} is strictly speaking a particle concept, while {\it fringes} have to do with waves. We are interested in how waves behave around a would-be-particle-caustic. 

After our paper was completed, we became aware of the work of
\cite{Banik:2017ygz}, who made the crucial observation that the Airy
function describes waves around a caustic. Our paper extends their
result in two ways: one is to derive the relevant wavefunction beyond
an energy eigenstate; the other is to test the analytic
prediction with simulations. We thank Elisa Todarello for pointing out
the work.


\section{The Airy solution for a linear potential}
\label{AiryAll}

Consider a bunch of particles moving in some gravitational potential well.
Their orbits run roughly radially outward from the center of the well, slowing
down in the process. Around apogee, where 
they spend the most time as the velocity drops to zero, the pile-up
of particles creates a spike in density. We wish to explore this
process in the wave regime.

We will do this in 3 steps. First, we will review the Airy function as
an energy eigenfunction of the Schr\"odinger equation with a linear
potential. The setup is appropriate for thinking about the vicinity of
apogee, where a particle gets slowed down to the point of
reversing motion. Second, we discuss how the Airy solution can be
adapted to describe wave behavior around a caustic close to apogee.
Here, the discussion is kept relatively brief, with the main
derivation relegated to Appendix \ref{HJ}. In these two steps, as a warm
up, we focus on waves in 1D (one spatial dimension). As a third step, we
discuss how to generalize to 3D, with the
restriction of spherical symmetry.
The idealization allows us to deduce simple expressions, which will be tested by more realistic simulations.

\subsection{The Airy eigenfunction---a review}
\label{Airy}
  
Taylor expanding the gravitational potential around apogee, it can be approximated
as linear plus a constant. Thus, the relevant Schr\"odinger equation for the
wavefunction $\psi$, in one spatial dimension, takes the form:
\begin{eqnarray} 
i \hbar \partial_t \psi = \left( - {\hbar^2 \over 2 m} \partial_x^2 + m V\right) \psi \, ,
\end{eqnarray}
with the gravitational potential approximated as
\begin{eqnarray}
\label{VaV0}
V = a x + V_0 \, ,
\end{eqnarray}
where $a$ is the local gravitational acceleration and $V_0$ is a constant.' This approximation will be accurate when the fringe separation is small compared to distance of the caustic radius, $r_c$.
Here $t$ and $x$ are the time and spatial coordinates.

A constant energy solution, with the time dependence $\psi \propto e^{-iE t}$,
satisfies
\begin{eqnarray}
\label{constantE}
\left( {E\over m} - V_0 \right) \psi = \left( - {\hbar^2 \over 2 m^2} \partial_x^2 + a x\right) \psi  \, .
\end{eqnarray}
It's convenient to choose $E/m - V_0 = 0$, or equivalently shift the origin such that $x=0$ is
the classical turning point (apogee). 
The physical solution, the one that decays in the forbidden region, is given by
the well known Airy function {\rm Ai}, up to an arbitrary overall normalization:
\begin{eqnarray}
\label{AiIntegral}
  \psi \propto {\rm Ai} \left( (2 m^2 a / \hbar^2)^{1/3} x \right) \, , \nonumber \\
  {\rm Ai} (z) \equiv {1\over 2\pi} \int_{-\infty}^\infty e^{i (z s + s^3/3)} ds \, .
\end{eqnarray}
In other words, ${\rm Ai}(z)$ satisfies: $(- \partial_z^2 + z) {\rm Ai}(z) = 0$.

\begin{figure}[!ht]
  \centering
  \includegraphics[width = .4\textwidth]{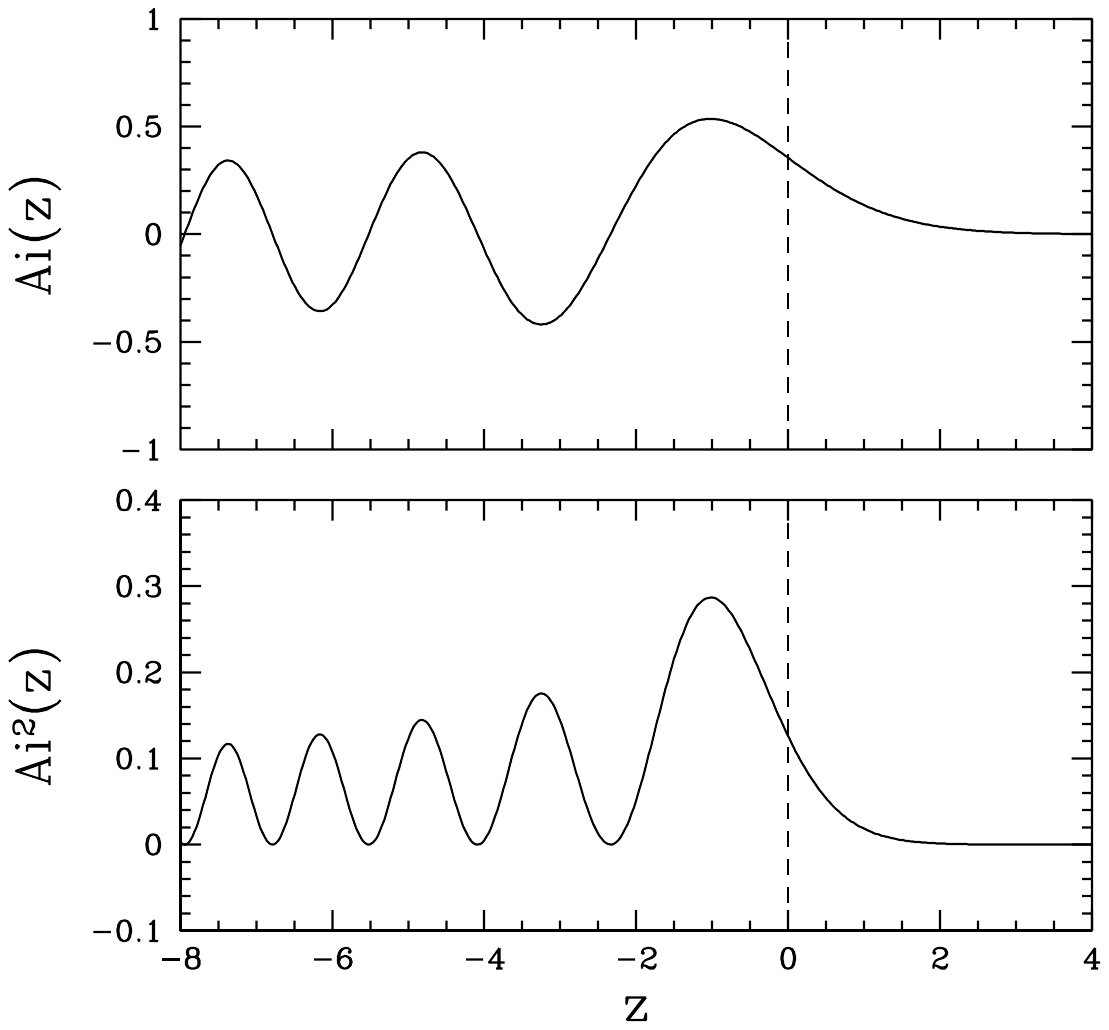}
  \vspace{-1.5cm}
	\caption{A plot of the Airy function ${\rm Ai} (z)$ (upper panel) and its square (lower panel).
          The Airy function solves the Schr\"odinger equation with a linear potential, with energy chosen
          such that $z=0$ corresponds to the turning point of a particle (dashed line).}
	\label{fig:Airy}
\end{figure}
      
A plot is shown in Figure \ref{fig:Airy}. Keep in mind the mass density is $\rho = m |\psi  (x)|^2$.
The wiggles in the Airy function thus translate into fringes in the dark matter density.
In particular, in terms of the density ($\propto {\rm Ai}^2$),
the first fringe (counting from the right) is separated from the second fringe by $\Delta z_{12}
\sim 2.2$.
This implies
\begin{equation}
\label{eqn:r_fringe}
\Delta x_{12} \sim 2.2 \left({\hbar^2   \over 2 m^2 a}\right)^{1/3} \, .
\end{equation}
We will make extensively use of this below.
We will approximate the width of the first fringe as
twice the distance between the first peak and the first zero of the
Airy function:
\begin{align} \label{eqn:caustic_width}
    \Delta x_\mathrm{c} \sim 2.6 \left( \frac{\hbar^2}{2 m^2 \, a} \right)^{1/3} \, .
\end{align}
In view of our intended application to dark matter
approaching apogee in a realistic gravitational potential, it should be kep in mind
that only the first few wiggles to the left of the turning-point (dashed line) should
be taken seriously. We expect the linear approximation to the potential to
receive significant corrections far from the turning-point.

Two comments are in order at this point.
First, $\psi$ here should be interpreted as a classical field
describing multiple particles, as opposed to
the standard textbook single-particle wavefunction (though we will
continue to refer to $\psi$ as the wavefunction).
In other words, we are interested
in a state in which there's high occupancy, for which $|\psi|^2$ can be interpreted as the
particle density (and hence the mass density $\rho = m |\psi|^2$).
A nice discussion can be found in \cite{Feynman}.
\footnote{The high occupancy means the quantum fluctuations are small, even as
  the classical density fluctuations, exemplified by $|\psi|^2$, can be large \cite{Guth2014, Eberhardt2023}.
  For our application, the appearance of $\hbar$ serves merely to convert mass $m$ into $m/\hbar$,
  which has the dimension of ${\rm length}^2/{\rm time}$. In other words, in combination with
  acceleration $a$, $(m^2 a / \hbar^2)^{-1/3}$ defines a length scale, which is our interference pattern scale.}
Second, the Airy solution is an energy eigenstate. It describes
a bunch of particles of the {\it same} energy. For practical applications,
such as splashback in clusters or tidal shells in merging galaxies, the situation might
not be exactly like this. We thus next turn to a brief discussion of
particle caustics---that is, the phenomenon of particle
pile-up---and establish a connection with the Airy solution for waves.

\subsection{The Airy solution for waves around a particle caustic}
\label{Caustics}


Let us first review what a caustic is. This is a concept that makes
sense for particles. (We will sometimes use the term {\it particle caustic}
to emphasize this fact.) Consider a bunch of particles, each on a trajectory as follows:
\begin{equation}
\label{qtox}
x(q,t) = q + v_0 (q) t - {1\over 2} a t^2 \, .
\end{equation}
Here, $q$ is the Lagrangian coordinate labeling the particle, $x$ is the (Eulerian) position
of the particle at time $t$, $v_0(q)$ is the initial velocity, and $a$ is the acceleration due to the
linear potential. Differentiating with respect to time, the corresponding velocity is
\begin{equation}
\label{qtov}
v(q, t) = v_0 (q) - a t \, .
\end{equation}

A caustic occurs where $\partial x / \partial q = 0$, i.e.
\begin{equation}
\label{causticCond}
1 + v_0'(q_c) t = 0 \, .
\end{equation}
Here $v_0'$ denotes $dv_0/dq$, and $q_c$ is the Lagrangian coordinate at which
the above condition is satisfied at the time of interest $t$. 
Let's call the corresponding Eulerian position $x_c$, and
velocity $v_c$:
\begin{equation}
\label{vstar}
v_c = v_0 (q_c) + {a \over v_0' (q_c)} \, .
\end{equation}
In general, the caustic need not coincide with the apogee, i.e.
$v_c$ need not vanish.


We are interested in what happens close to the caustic, at the same moment in time
$t$. We will mostly keep $t$ implicit. Taylor expanding
equation (\ref{qtox}), we see that
\begin{equation}
\label{dxCaustic}
x - x_c = - {v''_0 (q_c) \over 2 v'_0(q_c)} (q - q_c)^2 \, .
\end{equation}
Note how the linear in $q-q_c$ term disappears by virtue of the caustic condition.
Similarly, Taylor expanding equation (\ref{qtov}) and using the
above, we find
\begin{equation}
\label{vvc}
v-v_c = \pm \left( {2 v_0' (q_c) {}^3 \over v_0'' (q_c)} \right)^{1/2} \left[ -(x-x_c)\right]^{1/2} \, .
\end{equation}
Here, for the sake of concreteness, we assume $v_0' (q_c)$ and $v_0'' (q_c)$ are negative, such that
$x-x_c$ is also. It's straightforward to rewrite this expression for other sign choices.
An important corollary is that $\partial v/\partial x$ diverges at
the caustic. An illustration in phase space is depicted in the left
panel of Figure \ref{fig:background} below.
Note also, by mass conservation, the density $\rho \propto 1/|\partial x/\partial q|$
diverges at the caustic.

The above is the standard description of a particle caustic and its vicinity.
What is its wave analog? A convenient way to pass from the particle to
wave description is to make use of the Hamilton-Jacobi formalism, and
deduce the wavefunction in the WKB limit using the classical action.
The derivation is a bit long and is presented in Appendix
\ref{HJ}. We are particularly interested in a caustic that is close to
apogee, i.e. the pile-up of particles as they slow down to stand
still in a gravitational potential. In that case, it can be shown
the corresponding wavefunction in the vicinity of the caustic is
approximately given by:
\begin{equation}
\label{psiAicaustic2}
\psi \sim {\rm Ai}\left( (2m^2 a /\hbar^2)^{1/3} (x - x_c) \right) \, .
\end{equation}
where $x_c$ is the location of the caustic.
\footnote{\label{aApprox}
Essentially, when the caustic is close to apogee,
the combination $v_0' {}^3/v_0''$ in equation (\ref{vvc}), which
has the dimension of ${\rm length}/{\rm time}^2$, is well approximated
by the local acceleration $a$.}

\subsection{Generalizing to 3D}
\label{123}

We are interested in generalizing the above discussion to 3 spatial
dimensions, in a situation that's still effectively 1D in the sense of
having spherical symmetry.
The analog of equation (\ref{constantE}) for 3D, with the left hand
side set to zero, is
\begin{equation}
0 = \left( - {\hbar^2 \over 2 m^2} {1\over r^2} \partial_r (r^2 \partial_r) + a
  (r - r_c) \right) \psi  \, .
\end{equation}
We have chosen a linear potential around $r_c$, the radius at
which the caustic is located.

This can be rewritten as
\begin{equation}
0 = \left( - {\hbar^2 \over 2 m^2} \partial_r^2  + a
  (r - r_c) \right) r \psi  \, .
\end{equation}
Thus, we expect the analog of equation (\ref{psiAicaustic2}) to be
\begin{equation}
\label{psiAicaustic3}
\psi \sim {1\over r} {\rm Ai}\left( (2m^2 a /\hbar^2)^{1/3} (r - r_c) \right) \, .
\end{equation}
It follows that the separation between the first two fringes take the
same form as in the 1D case:
\begin{equation}
\label{eqn:r_fringe2}
\Delta r_{12} \sim 2.2 \left({\hbar^2   r_c^2 \over 2 m^2 GM_h }\right)^{1/3} \, ,
\end{equation}
where we have replaced the acceleration $a$ by $GM_h/r_c^2$
with $M_h$ being the mass enclosed within the radius $r_c$.
When applied to data, the precise caustic radius $r_c$ might be
unknown; one can use the location of the first fringe as a proxy.
\footnote{Strictly speaking, the 3D fringe separation should
  be slightly different from the 1D fringe separation, since
  the former has to do with peaks in ${\rm Ai}^2/r^2$
  while the latter has to do with peaks in ${\rm Ai}^2$.
  It can be shown the difference is negligible when
  the fringe separation is small compared to $r_c$.
  }

Recall that the de Broglie scale of a halo is $\lambda_{\rm dB} \equiv
\hbar/m\sigma$, where $\sigma$ is the velocity dispersion of the halo.
Approximating $\sigma^2 \sim GM_h / r_c$, we see that the fringe separation
can be rewritten as
\begin{equation}
\label{eqn:r_fringe3}
\Delta r_{12} \sim 1.75 \, \lambda_{\rm dB} \left( {r_c \over \lambda_{\rm
      dB}} \right)^{1/3} \, .
\end{equation}
Since the caustics we are interested in are on the outskirt
of a halo (recall they are associated with apogee), $r_c / \lambda_{\rm dB}$ is 
typically a large number. Thus, the fringe separation is enhanced
compared to the de Broglie scale based on the average halo velocity
dispersion. As we will see, an enhancement of an order of magnitude or more is
not unusual.

\section{Numerical simulations}
\label{NS}

In this section, we describe a series of numerical simulations,
first in 1D then in 3D, which will test the ideas presented in the
last section, in particular equations (\ref{psiAicaustic2}) and
(\ref{psiAicaustic3}).
These are simulations in which we solve the full
Schr\"odinger-Poisson system of equations without making any approximation.
In other words, we solve
\begin{eqnarray}
&& i \hbar \partial_t \psi = \left( - {\hbar^2 \over 2 m} \nabla^2 + m
  V\right) \psi \, , \nonumber \\
&& \quad \quad \nabla^2 V = 4 \pi G m |\psi|^2 \, ,
\end{eqnarray}
without making any approximation to $V$.

In parallel, we present N-body simulation results, with the initial
conditions chosen to match those of the wave simulations.
This will allow us to compare and constrast the predictions of
particles versus waves.

Below, we describe the initial conditions, the numerical solver
and the simulation results.

\subsection{Initial conditions}

Throughout this section we will refer to the box length as $L$, 
$M_\mathrm{tot}$ is the total mass, and $M_\mathrm{tot}/m$ is the the
squared norm of the wavefunction.

\textbf{One spatial dimension.} We simulate the gravitational collapse
of an initial spatial overdensity in a single dimension. The initial
wavefunction is given by
\begin{align}
    \psi(x) = \left( 1 + 0.1 \cos(2 \pi x / L) \right) /
  \sqrt{\mathrm{Norm}} \, .
\end{align}
The system has periodic boundary conditions, though we have checked
our results are not sensitive to this choice.

The corresponding initial condition for the N-body simulation is
chosen to match the density $\rho (x) = m |\psi(x)|^2$.
Since each N-body particle has the same mass, the N-body particles
are placed such that their density tracks $\rho (x)$.
A convenient way to enforce this is to first compute the cumulative
distribution function (CDF): 
\begin{align}
    {\rm CDF}(x) \equiv \frac{1}{M_{\rm tot}} \int_{-L/2}^{x} \rho (x') dx' \,.
\end{align}
The position for the $j$-th particle can be found by requiring
\begin{align}
    {\rm CDF}(x)|_{x=x_j} = \frac{j+1}{N_{\rm part}+2} \, ,
\end{align}
where $N_{\rm part}$ is the total number of particles.
The initial N-body particle velocities are zero, consistent with
the fact that the initial wavefunction is real.

\textbf{Three spatial dimensions.}
We carry out two kinds of simulations in 3D. One has spherical
symmetry, simulating the collapse of an initial Gaussian density
profile. The initial conditions are chosen in a manner analogous to
the 1D case. The second kind of simulations does not have spherical symmetry.
Instead, we simulate the growth of a halo that results from the
collision of many initial Gaussian
overdensities. This method is often used in the literature to
approximate the formation of halos. The initial
wavefunction takes the form
\begin{align}
    \psi(\vec x) = \sum_i^{N_g} e^{-(\vec x - \mu_i)^2 / 2 \sigma_x^2}
  / \sqrt{\mathrm{Norm}} \, ,
\end{align}
where $N_g$ is the number of Gaussians used and, $\mu_i$ is a random
position chosen uniformly in the box for each Gaussian, $\sigma_x$ is
the width of each Gaussian. We simulate systems with periodic and
non-periodic boundary conditions, and check that our results are not sensitive to this choice. 
The initial conditions are chosen such that new mass would fall into
the halo in discrete ``packages" and therefore be easier to
track. Also, adding asymmetry to the system lets us test our
prediction in a more realistic setting.

\begin{figure*}[!ht]
	\includegraphics[width = .97\textwidth]{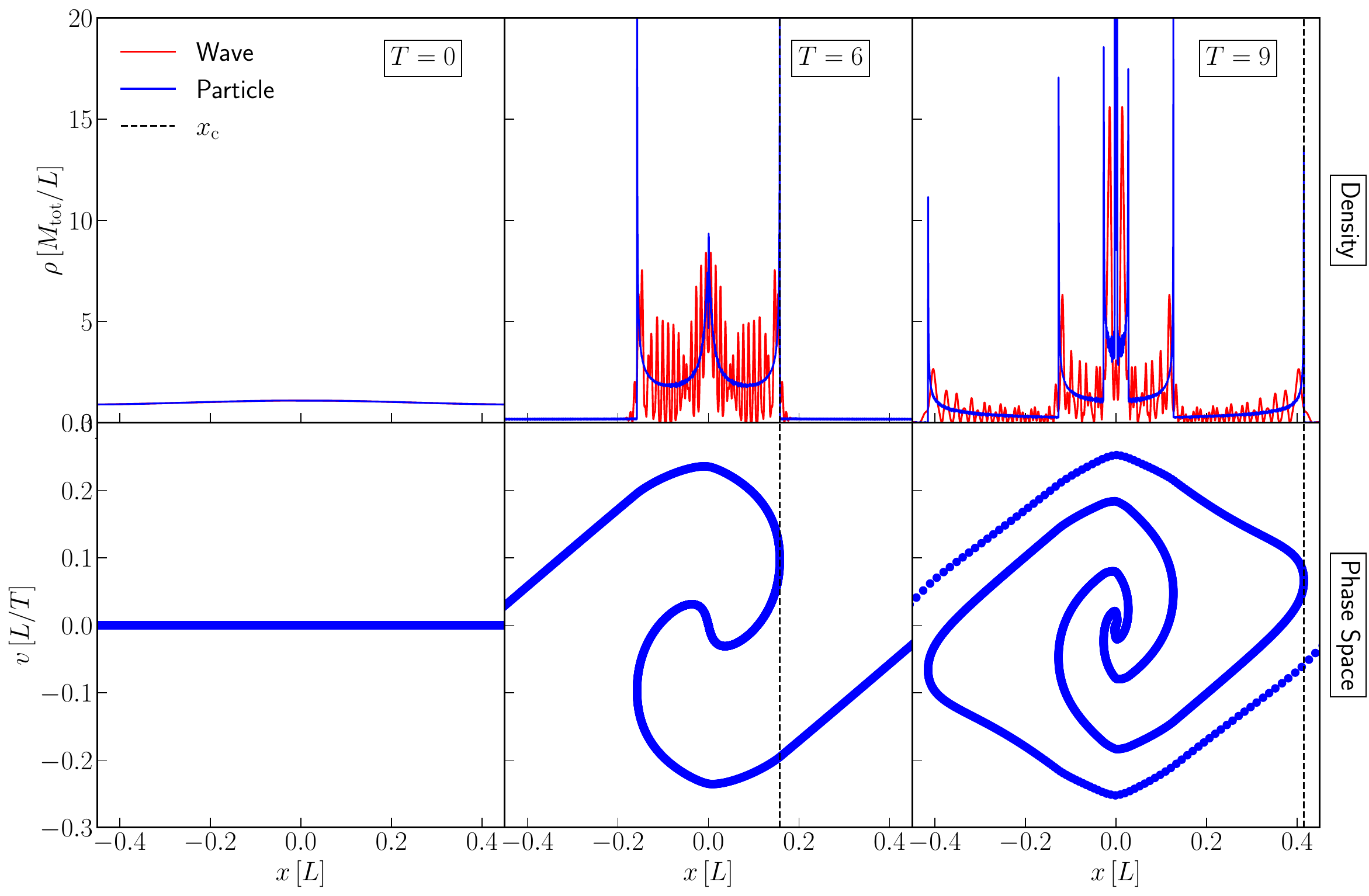}
	\caption{
          A simulation of the collapse of an initial
          overdensity in a single spatial dimension. Each column
          corresponds to a different snapshot at time $T$. \textbf{Top
            row.} We plot the density of a wave dark matter simulation
          (red) and an N-body particle simulation (blue). Broadly
          speaking, the two evolve similarly, but the
          wave dark matter simulation has interference
          fringes. \textbf{Bottom row.} We show the corresponding particle
          phase space at each snap shot. In the second and third columns,
          the dashed black line indicates the location of a caustic.
          Note how the velocity at the caustic nearly vanishes,
          i.e. it is close to apogee. Here $\hbar/m = 4 \times 10^{-4}$. All quantities, $\rho, v, x, T$, are
          shown in dimensionless code units.
        }
	\label{fig:evolution1D}
\end{figure*}

\begin{widetext}

\subsection{Numerical Solver}

Simulations are run using a standard pseudo-spectral leap frog solver
for wave dark matter simulations and a kick-drift-kick leap frog
solver for the N-body simulations. See \cite{Eberhardt2020} for an overview. 

For wave dark matter update this is written
\begin{align} \label{eqn:pseudoSpecUpdate}
    \psi_1(\vec x) &= e^{-i \delta t \, m V(\vec x,t) / \hbar / 2} \psi(\vec x,t) \, \, \, \mathrm{\textbf{[half-step kick update]}}  , \\
    \psi_2(\vec p) &= e^{-i \delta t \frac{\vec p^2}{2m} / \hbar} \mathcal{F}\left[ \psi_1(\vec x) \right](\vec p) \, \, \, \mathrm{\textbf{[full-step drift update]}} , \\
    \psi(\vec x, t + \delta t) &= e^{-i \delta t \, m V(\vec x,t + \delta t) / \hbar / 2} \mathcal{F}^{-1}\left[ \psi_2(\vec p) \right](\vec x) \, \, \, \mathrm{\textbf{[half-step kick update]}} ,
\end{align}
where $\mathcal{F}$ is the Fourier transform, and $\delta t$ is the
timestep.

And for the N-body update this is written
\begin{align}
    &\vec v_1 = \vec v(t) + \nabla V(\vec x,t) \, \delta t / 2 \, \, \, \mathrm{\textbf{[half-step kick update]}}  , \\
    &\vec r(t+\delta t) = \vec r(t) + \vec v_1 \, \delta t \, \, \, \mathrm{\textbf{[full-step drift update]}} , \\
    &\vec v(t+\delta t) = \vec v_1(t) + \nabla V(\vec x,t + \delta t) \, \delta t / 2 \, \, \, \mathrm{\textbf{[half-step kick update]}} .
\end{align}
The potential is calculated using the spectral method, i.e.,

\end{widetext}
\begin{align}\label{eqn:Poisson_k_space}
   V(\vec x,t) = \mathcal{F}^{-1}\left[ 4 \pi G \,  \frac{\mathcal{F}
    [\rho(\vec x',t)] (\vec k)}{\vec k^2} \right](\vec x) \, .
\end{align}

\begin{figure*}[!ht]
	\includegraphics[width = .66\textwidth]{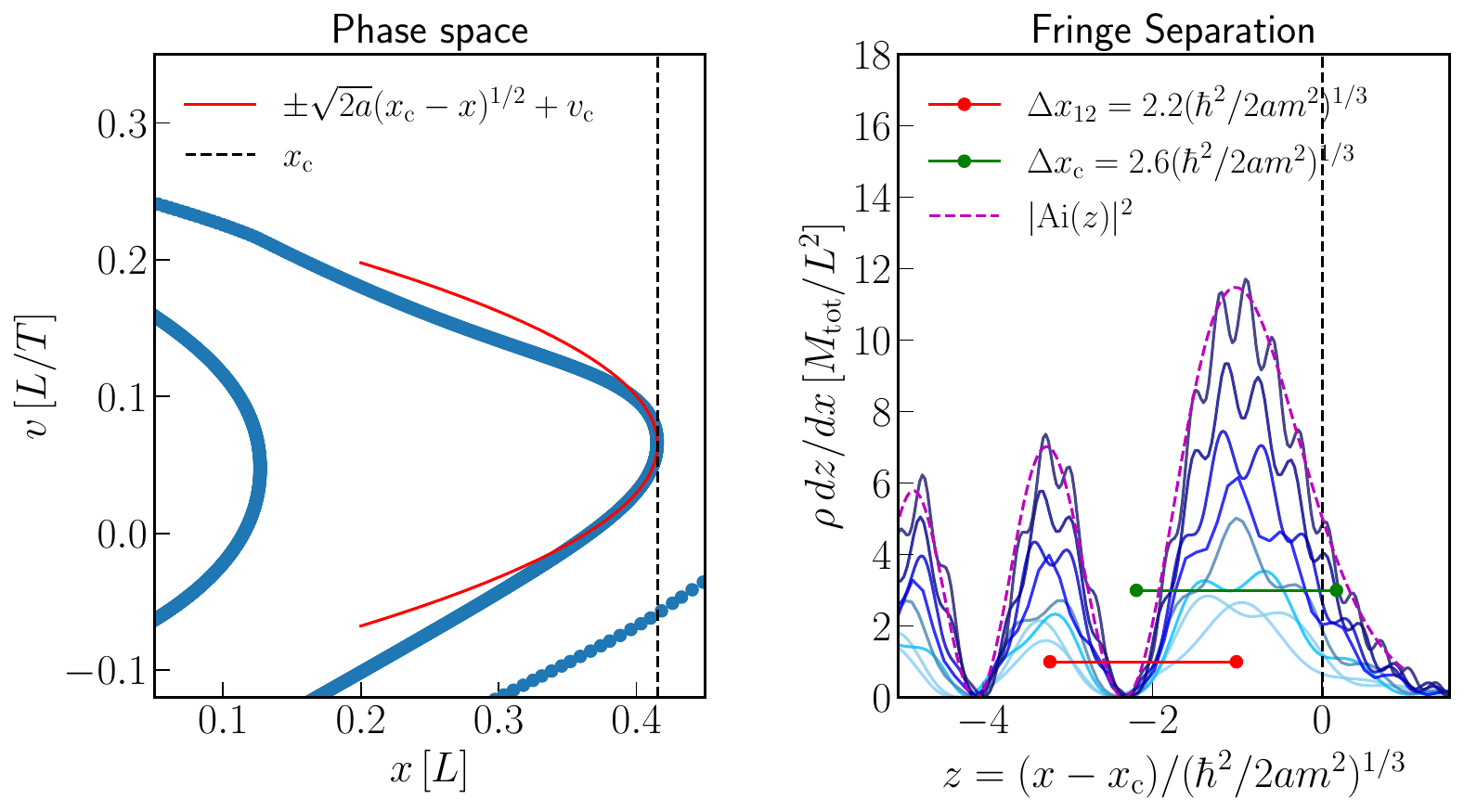}
	\caption{
          Here we zoom in on the caustic depicted in Figure \ref{fig:evolution1D}.
          In both panels the caustic location is labeled with a
          vertical dashed black line. \textbf{Left.} A plot of the
          particle phase space (blue) describing the collapse of an
          over-density due to gravity in a single spatial
          dimension. The red line indicates an approximation to the phase
          space in the vicinity of the caustic (equation \ref{vvc}).
          The acceleration $a$ is the local gravitational acceleration
          at the caustic.
          \textbf{Right.} A plot of the wave dark matter density
          (suitably normalized) in
          the vicinity of the caustic, for a series of wave simulations
          with the mass $m$ varying by a factor of $128$. 
          Each colored solid line represents a different $m$.
          The normalization of the y-axis is chosen so the smallest $m$ is at the
          bottom (lightest blue) and the largest $m$ is at the top
          (darkest blue).
          The Airy function, suitably normalized,
          provides a good match to all of them.
          To illustrate, we show with a dashed line
          an example that matches the top solid curve (modulo small
          wiggles; see text). The Airy function also gives the correct
          prediction for the fringe separation (red) and fringe width
          (green), in accordance with equations (\ref{eqn:r_fringe})
          and (\ref{eqn:caustic_width}).
                   }
	\label{fig:background}
      \end{figure*}

\begin{figure*}[!ht]
	\includegraphics[width = .7\textwidth]{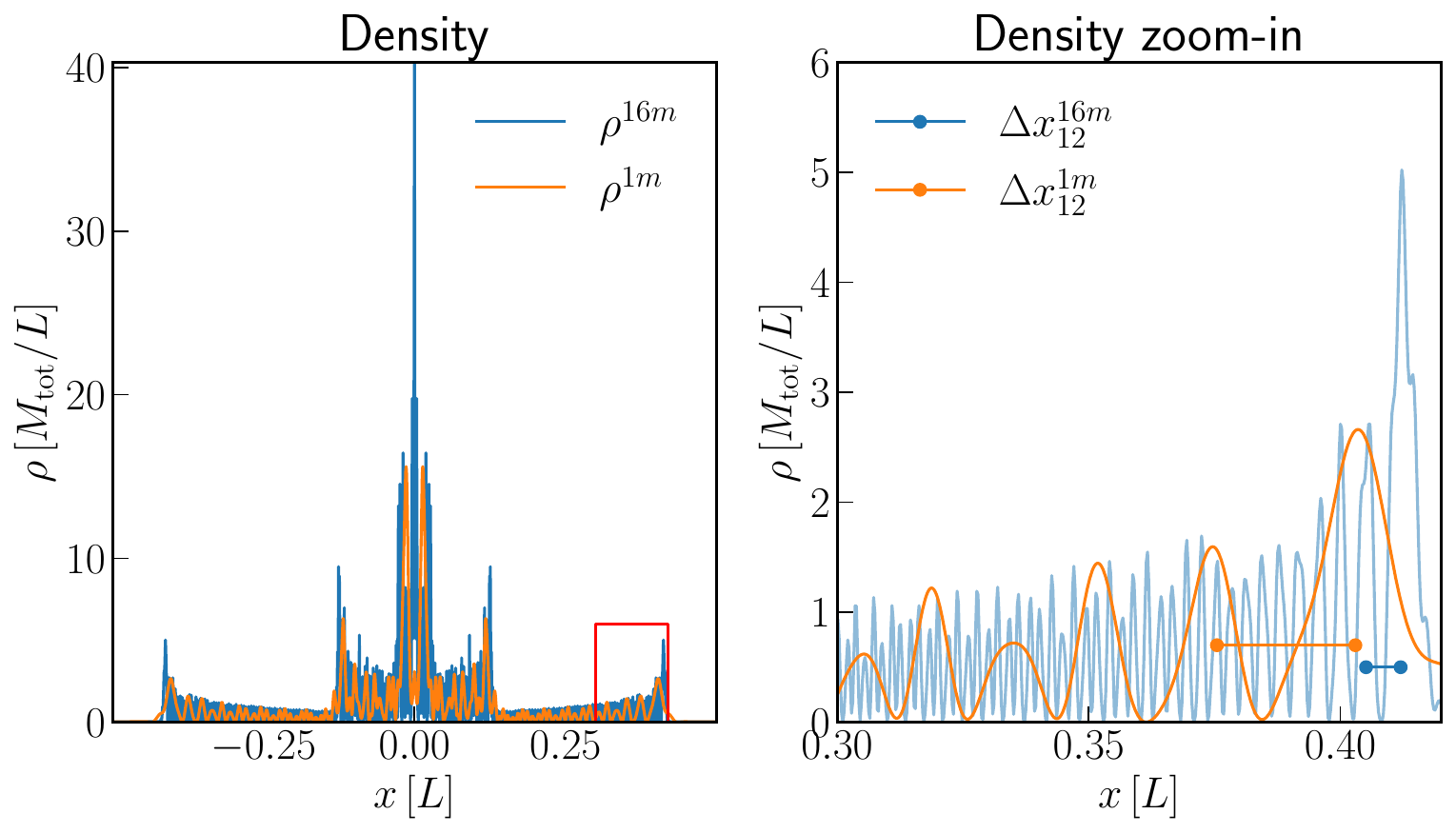}
	\caption{
          Two 1D wave simulations of the gravitational collapse of
          an initial overdensity for two different masses differing by
          a factor of $16$. The blue/orange line indicates the density
          profile for the higher/lower mass.
          The left panel shows the overall density profile in the
          simulation box. The right panel provides a zoom-in of the
          red region in the left panel. The horizontal lines indicate
          the respective expected fringe
          separation given by equation \eqref{eqn:r_fringe}. It
          provides an accurate description of the simulation results.}
	\label{fig:results1D}
      \end{figure*}
      
\subsection{Results} \label{sec:results}


\textbf{1D simulations.} We run a series of wave simulations of the collapse of an
initial overdensity in a single spatial dimension for $8$ different
masses spanning a factor of $128$. The value of the masses is given in terms of $\hbar/m$ which takes values $\hbar/m = 4 \times 10^{-4} / 2^n$ for $n = [0,1,\dots,7]$ in simulation units. In addition, we run an N-body particle
simulation corresponding to the same initial condition. Snapshots of both
(for one particular mass for the wave simulation) are shown in Figure
\ref{fig:evolution1D}. One can see how the density grows with time in
both the wave and particle simulations, with interference fringes
developing in the former. In particular, at the caustic (indicated by
the dashed line), the particle density becomes very large, while the
density remains finite in the wave simulation. 

In Figure \ref{fig:background}, we zoom in around the caustic of
interest. The left panel shows the phase space in the N-body
particle simulation after significant collapse has occurred.
The right panel shows the (suitably normalized) density for
the wave simulations of 8 different values of $m$. The Airy function
provides a good description of all of them, in accordance with
equation (\ref{psiAicaustic2}). In particular, both the separation
between the first two fringes, and the first fringe width, are
accurately described by equations (\ref{eqn:r_fringe}) and (\ref{eqn:caustic_width}). 
This is a non-trivial check: the numerical simulations make no
approximation about the potential, i.e. it's determined
self-consistently by solving the Poisson equation, while the analytic prediction based
on the Airy function assumes a constant local acceleration $a$ (here
measured from the simulations at the caustic location). The scaling of
the fringe pattern with $a$ and the mass $m$ works as predicted.

We note that the density in some of the simulations has additional
smaller scale oscillations on top of the main fringe structure. This
is due to the mass falling into the system for the first time that has
not yet reached apogee. In the N-body simulation, particles
corresponding to this kind of mass can be seen in the lower branch in
the phase space plotted in the left panel of Figure
\ref{fig:background}. The scale of these oscillations goes as the $\sim \hbar / m \Delta v$ where $\Delta v$ is the difference between the outward velocity of the caustic structure and the inward velocity of the accreting matter; we note that this is closer to but not equal to the de Broglie scale. These smaller scale oscillations and the
corrections to the constant acceleration approximation both provide
corrections to the true fringe separation. However, we can see our
approximation provides excellent agreement in the
1D systems tested. 

Figure \ref{fig:results1D} gives the global and zoomed-in view of the
density profile for two particular wave simulations in which the mass
differs by a factor of $16$. Here, we do not rescale the x-axis by the
characteristic scale $(\hbar^2/[2 a m^2])^{1/3}$, so one can see
explicitly that the fringe separation is larger when $m$ is smaller,
scaling in a way that matches our analytic prediction, equation
(\ref{eqn:r_fringe}). It is also worth pointing out that as
$m$ is lowered, the location of the first fringe moves further away
from the particle caustic, a behavior that is well matched by
the Airy function (\ref{psiAicaustic2}). 

\begin{figure*}[!ht]
	\includegraphics[width = .97\textwidth]{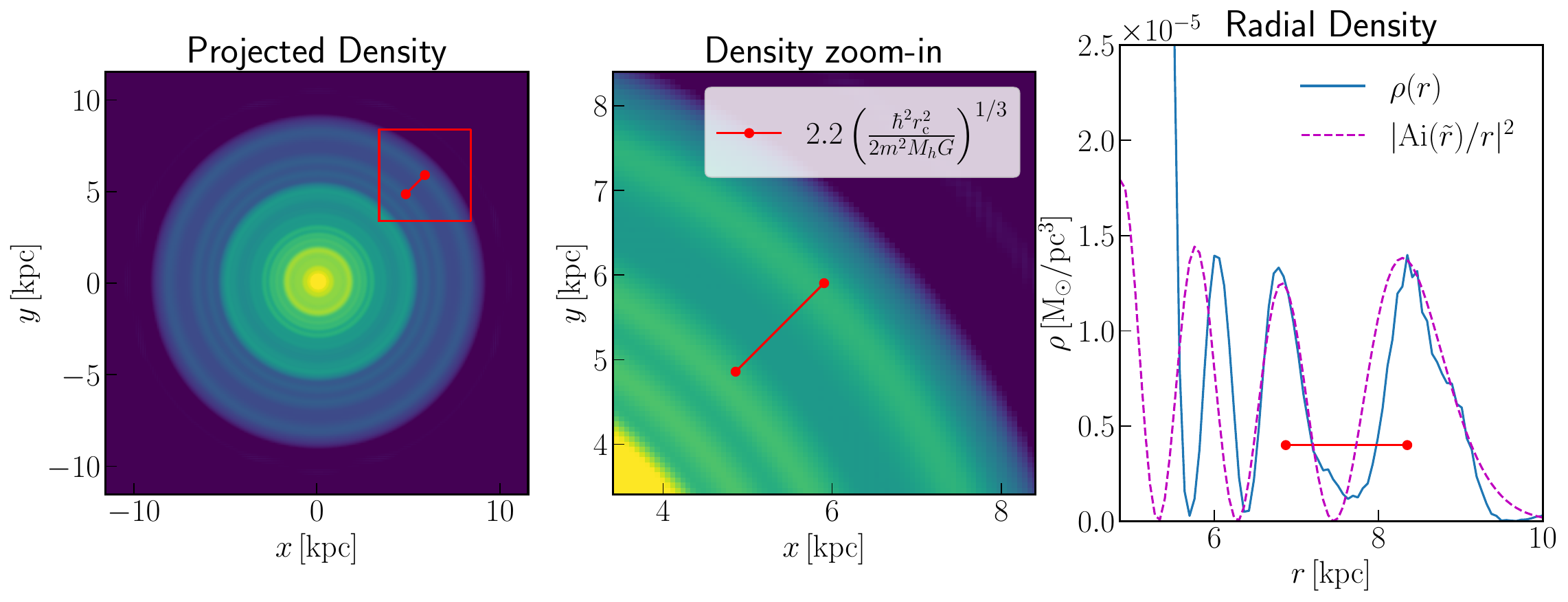}
	\caption{
          A 3D wave simulation of a spherically symmetric
          Gaussian blob undergoing gravitational collapse.
          \textbf{Left.} We show the projected log
          density. The collapse is spherically symmetric so we can see
          concentric rings from shells of materials and fringe
          structures. \textbf{Center.} We zoom in on the region with
          fringes (shown outlined in red in the left panel). We plot
          the expected separation in red superimposed on the
          fringes. We can see that the separation predicted in
          equation \eqref{eqn:r_fringe2} accurately describes the
          separation of the fringes. Here, the acceleration is
          $a=GM_h/r_c^2$, where $r_c$ is the caustic radius and $M_h$
          is the mass enclosed.
          \textbf{Right.} We plot the
          simulated density in blue and the Airy function prediction
          (equation \ref{psiAicaustic3}) in magenta. Here $\tilde r
          \equiv  (2m^2 GM_h / \hbar^2 r_c^2)^{1/3} (r - r_c)$, where
          $r_c$ is the caustic radius. 
          We can see that the Airy function describes the location and
          shape of the leading two fringes well but becomes inaccurate
          after this. In this simulation the total mass is
          $M_\mathrm{tot} = 4 \times 10^8 \, M_\odot$, the field mass
          $m = 5 \times 10^{-22} \, \mathrm{eV}$, and the grid is $384^3$. 
          }
	\label{fig:results3D_gauss}
\end{figure*}

\begin{figure*}[!ht]
	\includegraphics[width = .7\textwidth]{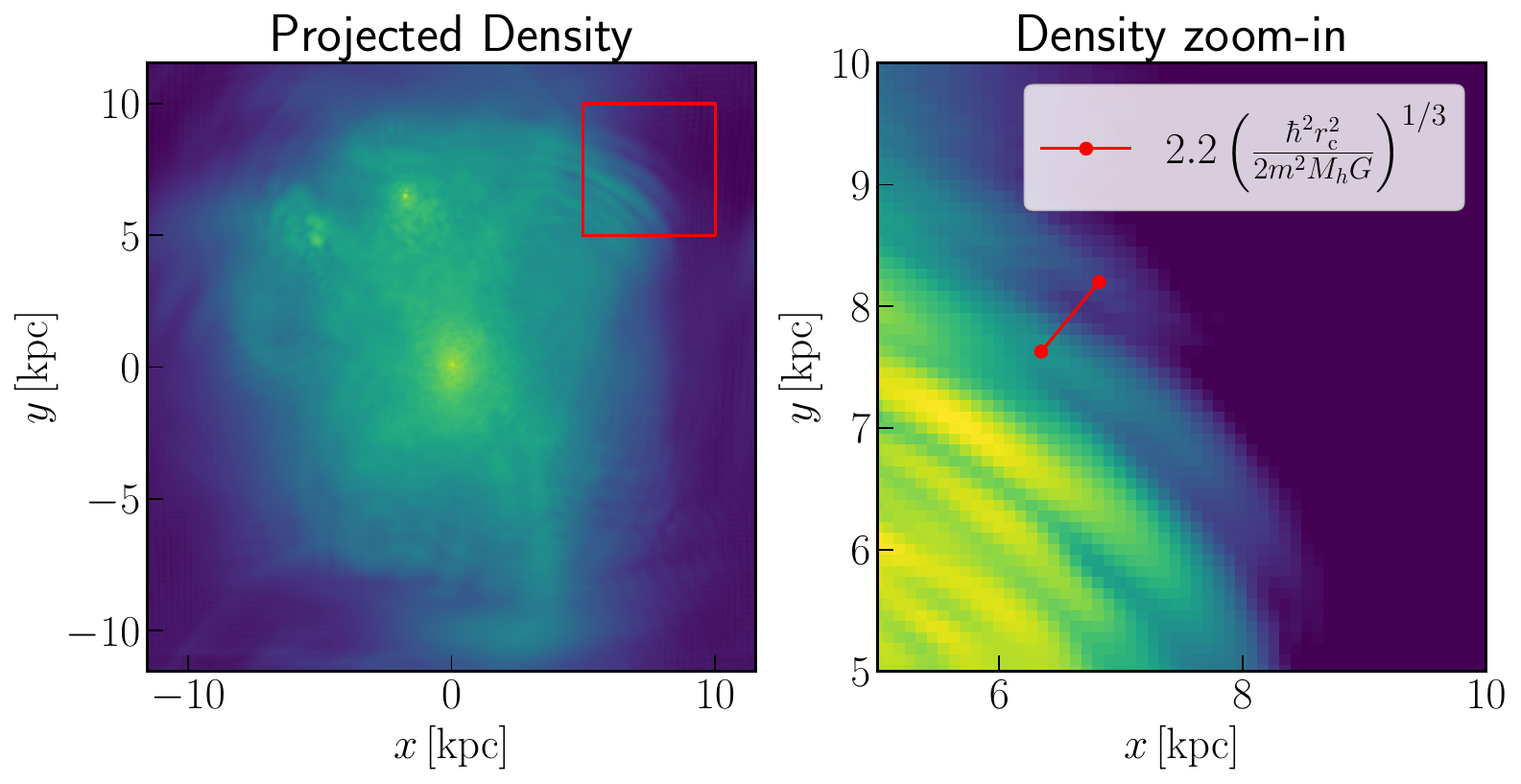}
	\caption{ A 3D wave simulation of Gaussian blobs collapsing and
          merging into a halo. \textbf{Left.} We show the projected
          log density. The soliton and de Broglie scale
          fluctuations are visible in the halo interior.
          In the periphery, we can see the fringes in
          the top right corner. These are associated with the most
          recent blob to have fallen into the halo. \textbf{Right.} We
          zoom in on the region with fringes (shown outlined in red in
          the left panel). We plot the expected separation in red
          superimposed on the fringes. We can see that the fringe
          separation predicted in equation \eqref{eqn:r_fringe2}
          matches what is seen in the simulation. In this simulation
          the total mass in the box is $M_\mathrm{tot} = 10^9 \,
          M_\odot$, and the field mass is $m = 10^{-21} \,
          \mathrm{eV}$. We use $N_g = 30$ equal mass Gaussian blobs,
          each with a width of $\sigma_x \approx 0.2 \, \mathrm{kpc}$,
          to construct the initial conditions. The grid is
          $256^3$. 
        }
	\label{fig:results3D}
\end{figure*}

\textbf{3D simulations.}
The first 3D wave simulation we run is the gravitational collapse of
a spherically symmetric Gaussian-shaped density.
The results are plotted in Figure \ref{fig:results3D_gauss}.
We can see that the first two fringes are well described by the Airy
solution, with the fringe separation matching the prediction from
equation \eqref{eqn:r_fringe2}. Subsequent fringes are not as well
matched by the Airy solution, presumably because the break down
of the constant acceleration approximation further away from the
caustic. Comparing against what we have seen in the 1D simulations,
the breakdown occurs sooner. This is not surprising, since the
acceleration goes roughly as $1/r^2$ in 3D , but constant in 1D,
at the outer edge of a halo.

As a test of the idea of Airy fringes in more realistic settings, we 
run 3D wave simulations of an ensemble of Gaussian blobs
collapsing and merging into a halo.
The blobs fall into the halo one at a time so that the apogee of the
most recently accreted matter is easy to locate.
This resembles the formation of tidal shells discussed in the
literature \cite{Malin1983,Sanderson2013,2015MNRAS.454.2472H,Pop2018,Kado-Fong2018,DongPaez2022}.
A snapshot of the resulting density field is shown in the left panel
of Figure \ref{fig:results3D}. The fringes associated with the most
recently accreted matter are clearly visible in the top right corner.
We zoom in on this particular region in the right panel of Figure
\ref{fig:results3D}, and find the separation of the fringes is well
described by equation \eqref{eqn:r_fringe2}, with the local
acceleration given by $a = GM_h/r_c^2$ , where $M_h$ is the current
halo mass enclosed within $r_c$.
We note that the fringe separation, $\Delta r_{12} \approx 0.744 \, \mathrm{kpc}$, has been enhanced by almost an order of magnitude over the halo de Broglie wavelength for the (visible as the average size of granules in the halo interior), $\lambda_\mathrm{db} = \hbar / m \sigma \sim 0.085 \, \mathrm{kpc}$. 




\section{Discussion}
\label{Discuss}

We have argued that the Airy function, which is based on
a linear potential approximation, provides an accurate description
of the interference pattern in wave dark matter in the vicinity of a
caustic. In particular, for a caustic close to apogee---the
pile-up as dark matter slows down to a stand-still---the first
and second fringes are separated by a predictable distance
(equation \ref{eqn:r_fringe2}):
\begin{equation}
\label{eqn:r_fringe2b}
\Delta r_{12} \sim 2.2 \left({\hbar^2  \over 2 m^2 a}\right)^{1/3} \, ,
\end{equation}
where the local acceleration $a$ can be estimated by $G M_h / r_c^2$,
with $r_c$ being the distance from the halo center and $M_h$ being the
mass enclosed.

We have put the above expression to the test in a number of ways:
by running wave simulations in 1D and 3D, in which the gravitational
potential is computed exactly (without the linear potential
approximation). We have even simulated situations in 3D that depart
from spherical symmetry. The above expression is found to be reliable
in all cases.

A remarkable corollary of this formula is that
$\Delta r_{12} \sim 1.75 \,\lambda_{\rm dB} (r_c / \lambda_{\rm
  dB})^{1/3}$ (equation \ref{eqn:r_fringe3}), where
$\lambda_{\rm dB}$ is the de Broglie wavelength $\hbar / m\sigma$
with $\sigma$ being the average velocity dispersion of the halo.
As such, the fringe separation is enhanced beyond the
halo de Broglie wavelength. Plugging in numbers:
\begin{eqnarray}
\label{Deltar12numbers}
  && \Delta r_{12} \sim 2.8 {\,\rm kpc}
  \left( {10^{-22} {\,\rm eV} \over m} \right)^{2/3} \nonumber \\
  && \quad \quad \quad \left( {100 {\,\rm km/s} \over \sigma} \right)^{2/3}
  \left( {r_c \over 100 {\,\rm kpc}} \right)^{1/3} \, .
\end{eqnarray}
The fringe separation to de Broglie ratio is
\begin{eqnarray}
  && {\Delta r_{12} \over \lambda_{\rm dB}} \sim 14.5
     \left( {m \over 10^{-22} {\,\rm eV}} \right)^{1/3} \nonumber \\
  && \quad \quad \quad \left( {\sigma \over 100 {\,\rm km/s}} \right)^{1/3}
  \left( {r_c \over 100 {\,\rm kpc}} \right)^{1/3}
\end{eqnarray}
The numbers chosen above are appropriate for a galaxy
(of velocity dispersion about $100$ km/s and size about $100$ kpc).
For a cluster of velocity dispersion about $1000$ km/s and size about
$1$ Mpc, the corresponding fringe separation is
$\Delta r_{12} \sim 1.3 {\,\rm kpc} \, (10^{-22} {\,\rm eV}/m)^{2/3}$,
and fringe separation to de Broglie ratio is $\Delta r_{12} / \lambda_{\rm dB}
\sim 67 \, (m / 10^{-22} {\,\rm eV})^{1/3}$ \, .

What are the prospects for observing such caustic fringes?
Recall that both splashback in the case of clusters, and tidal shells
in the case of galaxies , have been observed
\cite{Chang2018,Kado-Fong2018}.
The question is whether features on kpc scale or smaller
can be observed. In the case of cluster splashback, the observational data used
include galaxy counts and weak gravitational lensing. Achieving kpc
scale resolution seems challenging for these methods. Tidal shells for
galaxies are observed in the stellar distribution. A key question is
whether the stars are expected to trace the dark matter and reveal
the underlying fringe pattern if it exists.
The time-scale for the interference pattern to change is roughly
$\Delta r_{12} / v_c$, where $v_c$ is the velocity of the caustic.
Since the caustic is close to apogee, the caustic velocity is roughly $v_c \sim \epsilon \sigma$,
where $\epsilon$ is a small number. Thus, the interference time scale
is roughly $6 \times 10^8$ years for $\epsilon \sim 0.05$ assuming
the galaxy scale numbers. The dynamical time for the stars is roughly
$10^9$ years. The closeness of these two time scales suggests
a more careful computation is required to determine to what extent
stars would track the underlying fringe pattern.
Assuming they do, future observations (such as by the Roman Space
Telescope) have the potential of measuring features down to
a scale of $0.03$ kpc, implying the potential to probe $m \lsim 10^{-19}
{\,\rm eV}$. We caution this is likely overly optimistic, since higher masses
translate into a shorter interference time-scale. 

It's worth noting that there already exist in the literature a variety of constraints on
wave dark matter, some of which purport to rule out masses less than
about $10^{-19}$ eV \cite{Irsic:2017yje, Marsh2019, Rogers:2020ltq, dalal2022, Zimmermann:2024xvd,
Nadler:2024ejs}.
On the other hand, there are also papers that suggest data support $m$
as low as $10^{-22}$ eV \cite{Broadhurst:2024ggk, Palencia:2025wjw}. Given the possibility that each method suffers from its own
systematics, we find it useful to introduce a new
methodology which could provide a check on existing ones.
As such, looking for caustic fringes is an interesting option.
We stress that the scaling exhibited in equation
(\ref{eqn:r_fringe2b}) offers an important check on any claimed
detection. Fringes detected in different systems should be compared
to make sure they respect the expected scaling with acceleration.
The scaling in equation (\ref{eqn:r_fringe2b}) can also be exploited
to combine and stack data from different galaxies to enhance detection
sensitivity. We hope to pursue related issues of detectability in the
near future.

\section*{Acknowledgments}

We would like to thank Elisa Ferreira, Doddy Marsh and Masahiro Takada for useful
discussions. This work is supported by World Premier International Research Center
Initiative (WPI Initiative), MEXT, Japan, and by the Department of
Energy (DE-SC011941). LH thanks Eugene Lim and King's College London, and Centro de Ciencias de Benasque for hospitality.

\appendix

\section{Wave behavior around a particle caustic in a linear potential}
\label{HJ}

In this appendix, we discuss the wave behavior around a particle
caustic. Our discussion can be considered a generalization of \cite{Gough2022},
adding a linear potential to their free particle treatment.

In section \ref{Caustics}, we review the standard description of a
particle caustic and its vicinity.
To deduce the wave analog, a natural way to pass from particle to wave is to recall
the Hamilton-Jacobi description, in which the action $S$ is evaluated
on the classical path, and is regarded as a function of the end-point
of the path, at position $x$ and time $t$ (with the starting-point at
position $q$ and time $0$) \cite{LL}. It can be shown that
\begin{equation}
\label{HJeqs}
{\partial S(x, t; q, 0) \over \partial x} = m v \quad , \quad
{\partial S(x, t; q, 0) \over \partial t} = - H \, ,
\end{equation}
where $H$ is the Hamiltonian as a function of momentum $\partial
S/\partial x$ and position $x$. The corresponding wavefunction, under
the WKB approximation, is:
\begin{eqnarray}
\label{psiStheta}
\psi(x, t) = \int dq \, e^{i (S(x, t; q, 0) + \theta(q))/\hbar} \, ,
\end{eqnarray}
where $\theta (q) \equiv \int dq \, m v_0 (q)$ (the integration constant
contributes an overall phase which can be ignored). 
This expression can be motivated as follows: the dominant $q$ that
contributes, by the stationary phase approximation,
should be the one corresponding to the classical path,
i.e. $\partial S(x, t; q, 0)/\partial q = - m v_0 (q)$. This is
the flip side of equation (\ref{HJeqs}): differentiating
with respect to the end-point gives the momentum there; differentiating
with respect to the starting-point gives the negative momentum here \cite{LL}.
\footnote{Another way to see equation (\ref{psiStheta}) is this:
$\psi(x, t) = \langle x, t | \psi \rangle$ where $| \psi \rangle$
represents some state of interest. This can be rewritten as:
$\psi(x, t) = \int dq \langle x, t | q, 0 \rangle \langle q, 0 | \psi
\rangle$. The first factor in the integrand is approximated by
$e^{i S(x,t;q,0)/\hbar}$ where $S$ is evaluated on the classical path from
$q$ at time $0$ to $x$ at time $t$. The second factor gives rise to
$e^{i\theta(q)/\hbar}$, assuming the initial perturbation in suitably
chosen Lagrangian coordinates corresponds to perturbing the momentum
or the velocity. 
}

Equation (\ref{psiStheta}) might be unfamiliar to some readers.
It is useful to recall a simple example where this is exact.
Consider a free particle (no potential), for which $S(x,t;q,0)$ is
simply $(x-q)^2/(2t)$. Suppose all particles move with momentum $k$
initially, i.e. $\theta(q) = kq$. It's straightforward to show that
equation (\ref{psiStheta}) implies $\psi(x, t) \propto
e^{ikx-ik^2t/(2m)}$, as expected for free particles of momentum $k$.

For our case at hand, plugging equations (\ref{VaV0}) and (\ref{qtov}) into
$S = \int dt (m v^2/2 - m V)$, and replacing $v_0(q)$ by
$(x-q)/t + at/2$ from equation (\ref{qtox}), we find
\begin{eqnarray}
&& S(x, t; q, 0) = {1\over 2} m {(x-q)^2\over t} - {1\over 2} m a t x
  \nonumber \\ && \quad \quad \quad \quad -
{1\over 2} m a t q  - {1\over 24} m a^2 t^3 - m V_0 t \, .
\end{eqnarray}

Let's call the exponent $F(x,t;q,0) \equiv S(x,t;q,0) + \theta(q)$.
We are interested in Taylor expanding $F$ around $q=q_c$ to
obtain an approximate $\psi(x,t)$ for $x$ close to $x_c$.
$F(x,t;q_c,0)$ would contribute to a mere phase for $\psi(x,t)$ which
we will ignore. The first order term is: $\partial_q F(x,t; q,0) |_{q_c} (q-q_c) = - m (x - x_c)
(q - q_c)/t$. The second order term vanishes by virtue of
equation (\ref{causticCond}). The third order term is:
$\partial_q^3 F(x,t; q,0) |_{q_c} (q-q_c)^3/6 = m v_0''(q_c) (q -
q_c)^3/6$. Thus,
\begin{equation}
\psi(x, t) \propto \int dq \, e^{- i \left( {m(x-x_c)\over t}  (q-q_c) +
    {1\over 6} m (- v_0''(q_c)) (q - q_c)^3\right)/\hbar} \, .
\end{equation}
This can be compared against equation (\ref{AiIntegral}),
and so we have
\begin{equation}
\label{psiAicaustic}
\psi \propto {\rm Ai}\left( (2m^2 \left[ {v_0'(q_c) ^3 \over
  v_0''(q_c)} \right] /\hbar^2)^{1/3} (x - x_c) \right) \, .
\end{equation}

This is a key result. It shows that
close to a (particle) caustic, the wavefunction takes the
form of an Airy function \cite{Gough2022}. 
(Incidentally, Airy introduced the function when studying optical caustics.)
Comparing this against
equation (\ref{AiIntegral}), we see that
$v_0' {}^3 / v_0''$ evaluated at the caustic plays the role of
acceleration $a$, though there is nothing in the discussion
so far that suggests one is close to the other.

We thus close this appendix by discussing under what condition
the local acceleration at the caustic $a$ is indeed a good
approximation to $v_0' {}^3 / v_0''$.
This turns out to hold if the caustic is close to apogee,
that is to say, at the caustic where $1 + v_0'(q_c) t = 0$
(equation \ref{causticCond}), the velocity $v_0(q_c) - at$ also
approximately vanishes (equation \ref{qtov}).
Heuristically, we see that the two conditions put together suggests
$a \sim v_0^2 / \delta q$, where $\delta q$ is the characteristic
scale associated with the variation of $v_0$. In that case,
$v_0' {}^3 / v_0'' \sim v_0^2 / \delta q \sim a$.
This can be made more precise by noting that
if $v_0 (q) = \sqrt{-2 a q}$ ($q=0$ is the classical turning
point of the linear potential), then the caustic condition and the apogee
condition can be satisfied at the same time for some value $q_c$.

At the same $q_c$, it can be checked that $v_0' {}^3 / v_0'' = a$
exactly. We conclude that for a caustic close to apogee,
\begin{equation}
\label{psiAicaustic2b}
\psi \sim {\rm Ai}\left( (2m^2 a /\hbar^2)^{1/3} (x - x_c) \right) \, .
\end{equation}
where $x_c$ is the location of the caustic.

\section{Generalizing to 2D}
\label{2D}

Here, we carry out the analog of the generalization in Section
\ref{123} to 2D with rotational symmetry, or equivalently 3D with cylindrical symmetry.
The relevant Schr\"odinger equation is
\begin{equation}
- {\hbar^2 \over 2 m^2} \left(\partial_r^2 + {1\over r} \partial_r \psi
\right) + (m V(r) - E) \psi = 0 \, ,
\end{equation}
where we are keeping the potential general for now.
This can be rewritten as
\begin{equation}
\left(\partial_r^2 + {1\over 4 r^2} - {2 m^2 \over \hbar^2} \left( V(r) -
  {E\over m} \right) \right) (r^{1/2} \psi) = 0 \, .
\end{equation}
Let's assume $r_c$ is the location of the turning-point, i.e.
$V(r_c) - E/m = 0$. Taylor expanding
${1/4 r^2} - (2 m^2 /\hbar^2) \left( V(r) -
  {E/m} \right)$ around $r_c$, we find
\begin{equation}
  \left(\partial_r^2 + {1\over 4 r_c^2} -  {2m^2 a \over \hbar^2}
    (1 + \gamma) (r-r_c) \right) (r^{1/2}\psi) = 0
\end{equation}
where $a \equiv V'(r_c)$.
\begin{equation}
\gamma \equiv {\hbar^2 \over 4 m^2 ar_c^3} \, .
\end{equation}
Thus, we find
\begin{equation}
\psi \propto r^{-1/2} {\,\rm Ai} \left[ \left({2m^2 a \over \hbar^2} (1 +
  \gamma) \right)^{1/3} \left( r - r_c {2 + 3\gamma \over 2 + 2\gamma}
\right) \right] \, .
\end{equation}
For practical applications, $\gamma$ is likely small.

\bibliography{BIB.bib}

\end{document}